\documentclass[a4paper]{article}

\usepackage{17lomcon}        
\usepackage{cite}             
\usepackage{epsfig}           

\usepackage{xspace}
\usepackage{floatrow}
\usepackage{amsmath}
\usepackage{amsfonts}
\usepackage{amssymb}
\usepackage{calrsfs}
\usepackage[colorlinks=true,allcolors=blue]{hyperref}

\newcommand{\epem}{e^+e^-}
\newcommand{\alphas}{\alpha_{\rm s}}
\newcommand{\sqrts}{\sqrt{\rm s}}
\newcommand{\qqbar}{q\overline{q}}
\newcommand{\bbbar}{b\overline{b}}
\newcommand{\ttbar}{t\overline{t}}
\newcommand{\mZ}{m_{_\mathrm{\rm Z}}}
\newcommand{\mW}{m_{_\mathrm{\rm W}}}

\newcommand*{\cm}{c.m.\@\xspace}
\newcommand*{\eg}{e.g.\@\xspace}

\def\ttt#1{\texttt{\small #1}}
\def\cO#1{{{\cal{O}}}\left(#1\right)}

\begin{document}



\title{PHYSICS AT THE FCC-ee}

\author{David d'Enterria\email{dde@cern.ch}}
\affiliation{CERN, EP Department, 1211 Geneva, Switzerland}


\date{15/02/2016}

\maketitle

\vspace{0.2cm}
\begin{abstract}
The physics program accessible in $\epem$ collisions at the Future Circular Collider (FCC-ee) is 
summarized. The FCC-ee aims at collecting multi-ab$^{-1}$ integrated luminosities in $\epem$ at $\sqrts$~=~90,
160, 240, and 350~GeV, yielding 10$^{12}$ Z bosons, 10$^{8}$ W$^+$W$^-$ pairs, 10$^{6}$ Higgs bosons and
$4\cdot 10^{5}$ top-quark pairs per year. Such huge data samples combined with a $\cO{100~\rm keV}$ \cm energy
uncertainty will allow for Standard Model measurements with unparalleled precision and searches for
new physics in regions not probed so far. The FCC-ee will be able to (i) indirectly discover 
new particles coupling to the Higgs and electroweak bosons up to scales $\Lambda \approx$~7 and 100~TeV;
(ii) perform competitive SUSY tests at the loop level in regions beyond the LHC reach; and 
(iii) achieve the best potential in direct collider searches for dark matter and sterile neutrinos with masses below 60~GeV.
\end{abstract}

\section{Introduction}

\begin{figure}[htbp!]
\centering
\includegraphics[scale=0.63]{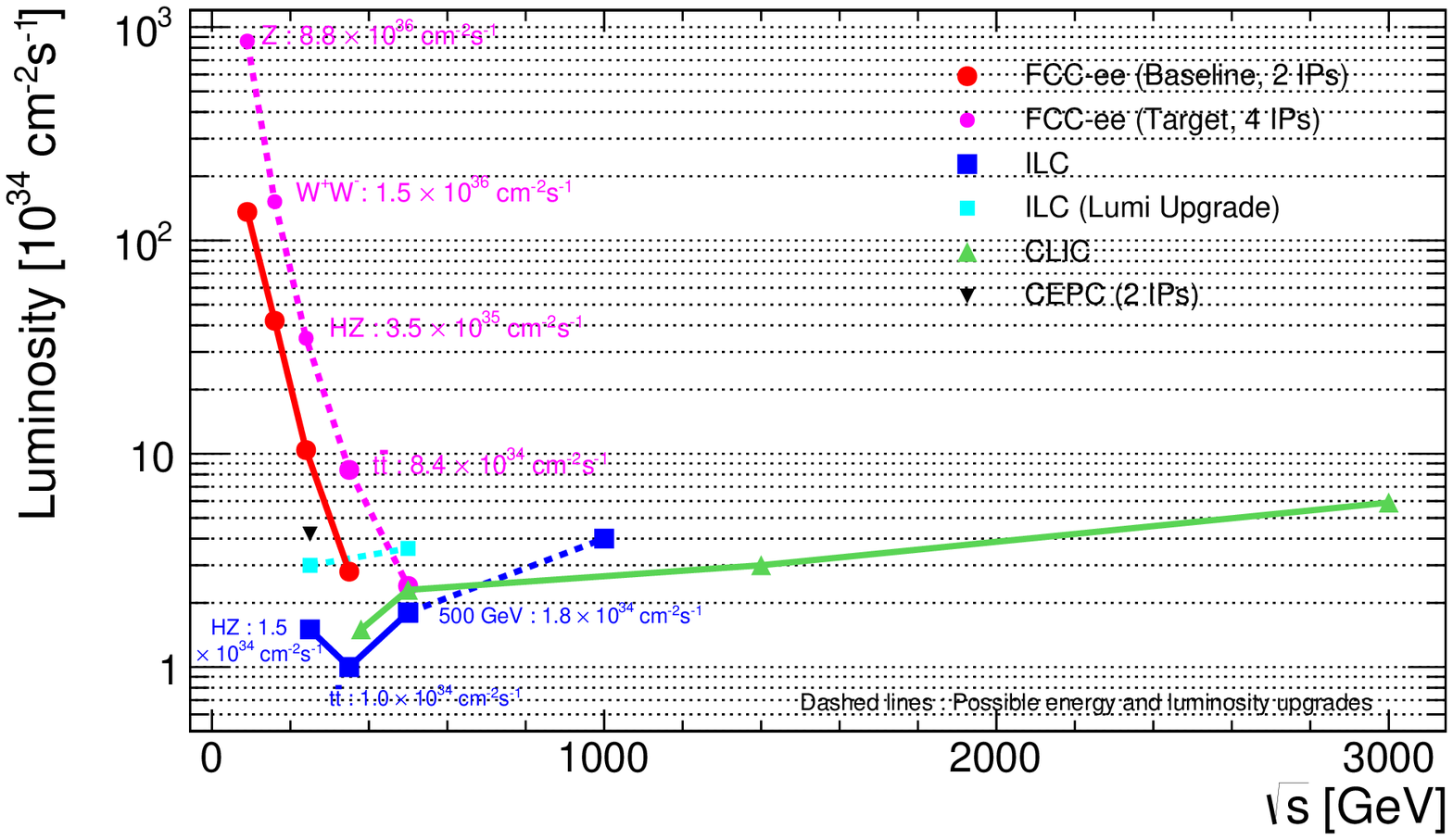}
\caption{Target luminosities as a function of center-of-mass energy, in the range
  $\sqrts\approx$~90--3000~GeV, for circular (FCC-ee, CEPC) and linear 
  (ILC, CLIC) $\epem$ colliders currently under consideration.} 
\label{fig:ee_lumi}
\end{figure}

Today, the understanding of particle physics is theoretically encoded in the Standard Model (SM), a renormalizable
quantum field theory --unifying quantum mechanics and special relativity-- that describes the fundamental
interactions (except gravity) via a local SU(3)$\times$ SU(2)$\times$U(1) 
gauge-symmetry group.
The three gauge-symmetry terms give rise to the strong, weak and electromagnetic forces, while the particles
fall into different representations of these symmetry groups. The SM Lagrangian (without neutrino masses)
contains 19 free parameters to be determined experimentally: 3 gauge couplings, 9 Higgs--fermion Yukawa couplings, 
3 mixing-angles, 2 Charge-Parity (CP) phases, and 2 Higgs boson couplings.
Despite its tremendous success to accurately and consistently describe all phenomena observed at particle
accelerators so far --including the recent experimental confirmation of the existence of its last missing
piece, the Higgs boson-- the theory is not complete and has several outstanding open questions to solve:
\begin{enumerate}
\item Dark matter (DM): The SM describes only $\sim$4\% of the universe energy budget, the rest being in
  unknown DM (and dark energy) forms, pointing to the existence of 
  new massive particles (such as \eg\ SUSY partners, axions, heavy $\nu$'s...).
\item Flavour problem: The huge dominance of matter over antimatter in the universe cannot be explained by the
  known SM sources of CP violation. More generally, the SM fails to explain the rationale behind the
  observed pattern of fermion families masses and flavour mixings. 
\item Neutrino masses: The generation of non-zero neutrino masses, called for by the observation of their flavour
  oscillations, is beyond the SM (BSM) and requires new particles such as right-handed $\nu$'s.
\item Hierarchy (or ``fine tuning'', ``naturalness'') problem: Quadratically-divergent virtual SM corrections
  affect the running of the Higgs boson mass between the widely separated electroweak and Planck scales,
  calling for new  (\eg\ supersymmetric) particles to stabilize such ``untamed'' quantum corrections.
\item Strong CP problem: The absence of CP-violation in QCD is naturally explained by postulating a
  (Peccei-Quinn) symmetry that gives rise to a new particle (axion).
\item Other fundamental SM issues include the current inability to explain dark energy, the cosmological constant, the
  origin of inflation, or gauge-gravity unification. 
\end{enumerate}
Many (or all) such fundamental questions may likely {\it not} be fully answered through the study of p-p
collisions at the 
LHC. Despite their lower center-of-mass energies, $\epem$ colliders feature several advantages compared to
their hadronic counterparts in terms of new physics studies: 
(i) direct model-independent searches for new particles coupling to Z*/$\gamma$* with masses up to
$m\approx\sqrts/2$;
(ii) signals and backgrounds with very precisely known theoretical QCD and electroweak corrections with
uncertainties (well) below 1\% in many cases; and
(iii) very clean experimental environment with final states free of ``holes'' or ``blind spots'' typical
of p-p searches involving new particles with difficult decay modes (\eg\ in certain SUSY scenarios with soft
final-states and/or without missing transverse energy).
Thus, combined with high-luminosities, an $\epem$ collider can provide access to studies with  permil-level
precision $\delta X$, allowing one to place indirect constraints on BSM physics --appearing as virtual
corrections to well-controlled processes-- up to very-high energies $\Lambda\propto$~(1~TeV)/$\sqrt{\delta
  X}$. Plans exist to build future circular (FCC-ee~\cite{FCCee}, CEPC~\cite{CEPC}) and/or linear
(ILC~\cite{ILC}, CLIC~\cite{CLIC}) $\epem$ colliders (Fig.~\ref{fig:ee_lumi}).

The advantages of circular over linear $\epem$ machines are (i) much larger luminosity below
$\sqrts\approx$~400~GeV (thanks to much higher collision rates, adding continuous top-up injection to
compensate for luminosity burnoff); (ii) the possibility to have several interaction points (IPs); and 
(iii) a very precise measurement of the beam energy E$_{\rm beam}$ through resonant transverse
depolarization~\cite{Koratzinos:2015hya}. Linear $\epem$ colliders, on the other hand, feature (i) much larger
$\sqrts$ reach (circular colliders are not competitive above $\sqrts\approx$~400~GeV due to synchroton
radiation scaling as E$_{\rm beam}^4/R$), and (ii) easier longitudinal beam polarization. At the FCC (with a
radius R~=~80--100~km), $\epem$ collisions present clear advantages with respect to the older LEP
($\times$10$^4$ more bunches, and $\delta E_{\rm beam}$~=~$\pm$0.1~MeV compared to $\pm$2~MeV) and the ILC
(crab-waist optics scheme, up to 4 IPs) yielding luminosities $\times$(10$^4$--10) larger than both machines
in the $\sqrts$~=~90--350~GeV range~\cite{Zimmermann}.\\

Table~\ref{tab:runs} lists the target FCC-ee luminosities and associated total number of events  at each $\sqrts$. 
They have been obtained for the relevant cross sections including initial state radiation and smearings due
to beam-energy spreads: $\sigma_{\rm \epem\to Z}=$~43~nb, $\sigma_{\rm \epem\to H}=$~0.29~fb, $\sigma_{\rm \epem\to W^+W^-}=$~4~pb, 
$\sigma_{\rm \epem\to HZ}=$~200~fb, $\sigma_{\rm \epem\to \ttbar}=$~0.5~pb, and $\sigma_{\rm \epem\to VV \to H}=$~30~fb.
The completion of the FCC-ee core physics program (described in the next sections) requires $\sim$10 years of
running. 

\renewcommand\arraystretch{1.25}
\begin{table}[htpb!]
\centering
\small
 \begin{tabular}{l|c|c|c|c|c|c} \hline 
 $\sqrts$ (GeV): & 90 (Z) & 125 (eeH) & 160 (WW) & 240 (HZ) & 350 ($\ttbar$) &  350 (WW$\to$H) \\ \hline\hline
 $\cal{L}/$IP (cm$^{-2}$\,s$^{-1}$) & 2.2$\cdot$10$^{36}$ & 1.1$\cdot$10$^{36}$ & 3.8$\cdot$10$^{35}$ & 8.7$\cdot$10$^{34}$ & 2.1$\cdot$10$^{34}$ & 2.1$\cdot$10$^{34}$ \\ 
 $\cal{L}_{\rm int}$ (ab$^{-1}$/yr/IP) & 22 &  11 & 3.8 & 0.87 & 0.21 & 0.21 \\ 
 Events/year (4 IPs) & 3.7$\cdot$10$^{12}$ & 1.2$\cdot$10$^{4}$ & 6.1$\cdot$10$^{7}$ & 7.0$\cdot$10$^{5}$ & 4.2$\cdot$10$^{5}$ & 2.5$\cdot$10$^{4}$ \\ 
 Years needed (4 IPs) &  2.5 &  1.5 & 1 & 3 & 0.5 & 3 \\ \hline
 \end{tabular}
 \caption{Target luminosities, events/year, and years needed to complete the W, Z, H and
   top-quark programs at FCC-ee. [Note that $\cal{L}$~=~10$^{35}$~cm$^{-2}$\,s$^{-1}$ corresponds to
$\cal{L}_{\rm int}$~=~1~ab$^{-1}$/yr for 1 yr = 10$^7$ s].}
 \label{tab:runs}
\end{table}

These proceedings present succinctly the FCC-ee physics program. More detailed information can be found
in~\cite{FCCee}, and in the dedicated physics FCC-ee (mini)workshops organized recently~\cite{fcc-ee}.

\section{Indirect  constraints on BSM via high-precision electroweak and top physics}
\label{sec:SM}

Among the main goals of the FCC-ee is to collect multi-ab$^{-1}$ at $\sqrts\approx$~91~GeV (Z pole), 160~GeV
(WW threshold), and 350~GeV ($\ttbar$ threshold) in order to measure key properties of the W and Z bosons 
and the top-quark, as well as other fundamental SM parameters, with unprecedented precision.
The combination of huge data samples available at each $\sqrts$ and the exquisite control of the \cm energy
(at the $\pm$100~keV level) leading to very accurate energy threshold scans, allows the experimental
precision of many SM parameters to be improved by a factor better than 25 with respect to the current
state of the art (Table~\ref{tab:SM})~\cite{Tenchini:2014lma}. Some FCC-ee experimental precision targets are
$\pm$100~keV for $\mZ$, $\pm$500~keV for $\mW$, $\pm$10~MeV for $m_\mathrm{t}$, a relative statistical
uncertainty of the order of 3$\cdot$10$^{-5}$ for the QED $\alpha$ coupling (through $\epem\to\mu^+\mu^-$
forward-backward asymmetries above and below the Z peak)~\cite{Janot:2015gjr}, one-permil for the QCD coupling
$\alphas$ (through hadronic Z and W decays)~\cite{d'Enterria:2015toz}, and one-permil on the electroweak top
couplings $F^{\gamma\,t,Z\,t}_\mathrm{1V,2V,1A}$ (through angular distributions in $\epem\to\ttbar\to 
\ell\nu\qqbar\bbbar$)~\cite{Janot:2015yza}. In many cases, the dominant uncertainty will be of theoretical
origin, and developments in the calculations are needed in order to match the expected experimental
uncertainty.\\

\renewcommand\arraystretch{1.25}
\begin{table}[htpb!]
\centering
\small
 \begin{tabular}{@{}c@{}|@{}c@{}|@{}c@{}|@{}c@{}|@{}c@{}|@{}c@{}} \hline 
 Observable & Measurement & Current precision & \;FCC-ee stat.\; & \;Possible syst.\; & Challenge \\ \hline\hline
 $\mZ$ (MeV) & Z lineshape & $91187.5 \pm 2.1$ & 0.005 & $<0.1$ & QED corr. \\ 
 $\Gamma_{_\mathrm{Z}}$ (MeV) & Z lineshape & $2495.2 \pm 2.3$ & 0.008 & $<0.1$ & QED corr. \\ 
 $R_\ell$ & Z peak & $20.767 \pm 0.025$ & 0.0001 & $<0.001$ & QED corr. \\ 
 $R_\mathrm{b}$ & Z peak & $0.21629 \pm 0.00066$ & 0.000003 & $<0.00006$ & $g\rightarrow \mathrm{b\bar{b}}$ \\ 
 $N_\nu$ & Z peak & $2.984 \pm 0.008$ & 0.00004 & $0.004$ & Lumi meas. \\ 
 $N_\nu$ & $\mathrm{e^+e^- \rightarrow \gamma\, Z(inv.)}$ & $2.92 \pm 0.05$ & 0.0008 & $<0.001$ & -- \\ 
 $A_{_\mathrm{FB}}^{\mu\mu}$ & Z peak & $0.0171 \pm 0.0010$ & 0.000004 & $<0.00001$ & $E_\mathrm{beam}$ meas. \\ 
 $\alphas(\mZ)$ & $R_\ell, \sigma_\mathrm{had}, \Gamma_{_\mathrm{Z}}$ & $0.1190 \pm 0.0025$ & 0.000001 &
 $0.00015$ & New physics \\ 
 $\;1/\alpha_{_\mathrm{QED}}(\mZ)\;$ & \;$A_{_\mathrm{FB}}^{\mu\mu}$ around Z peak\; & $128.952 \pm 0.014$ & 0.004 & 0.002 & EW corr. \\ \hline
 $\mW$ (MeV) & WW threshold\ scan & $80385 \pm 15$ & 0.3 & $<1$ & QED corr. \\ 
 $\alphas(\mW)$ & $\Gamma_{_{\rm W}}, B^{^{\rm W}}_\mathrm{had}$ & $B^{^{\rm W}}_\mathrm{had} = 67.41 \pm 0.27$ & 0.00018 & $0.00015$ & \;CKM matrix\;\\ \hline
 $m_\mathrm{t}$ (MeV)& $\ttbar$ threshold scan & $173200 \pm 900$ & 10 & 10& QCD \\ 
 $\Gamma_\mathrm{t}$ (MeV)& $\ttbar$ threshold scan & $1410^{+290}_{-150}$ & 12 & ? &  $\alphas(\mZ)$  \\ 
 $y_\mathrm{t}$ &  $\ttbar$ threshold scan & $\mu = 2.5 \pm 1.05$ & 13\% & ? &  $\alphas(\mZ)$  \\ 
 $F^{\gamma\,t,Z\,t}_\mathrm{1V,2V,1A}$& d$\sigma^{\ttbar}$/dx\,d$\cos(\theta)$ & \;4\%--20\% (LHC-14 TeV)\; & (0.1--2.2)\% & (0.01--100)\% & -- \\ \hline
 \end{tabular}
 \caption{Examples of achievable precisions in representative Z, W and top measurements at FCC-ee.}
 \label{tab:SM}
\end{table}

\begin{figure}[htbp!]
\centering
\includegraphics[width=0.49\columnwidth,height=4.5cm]{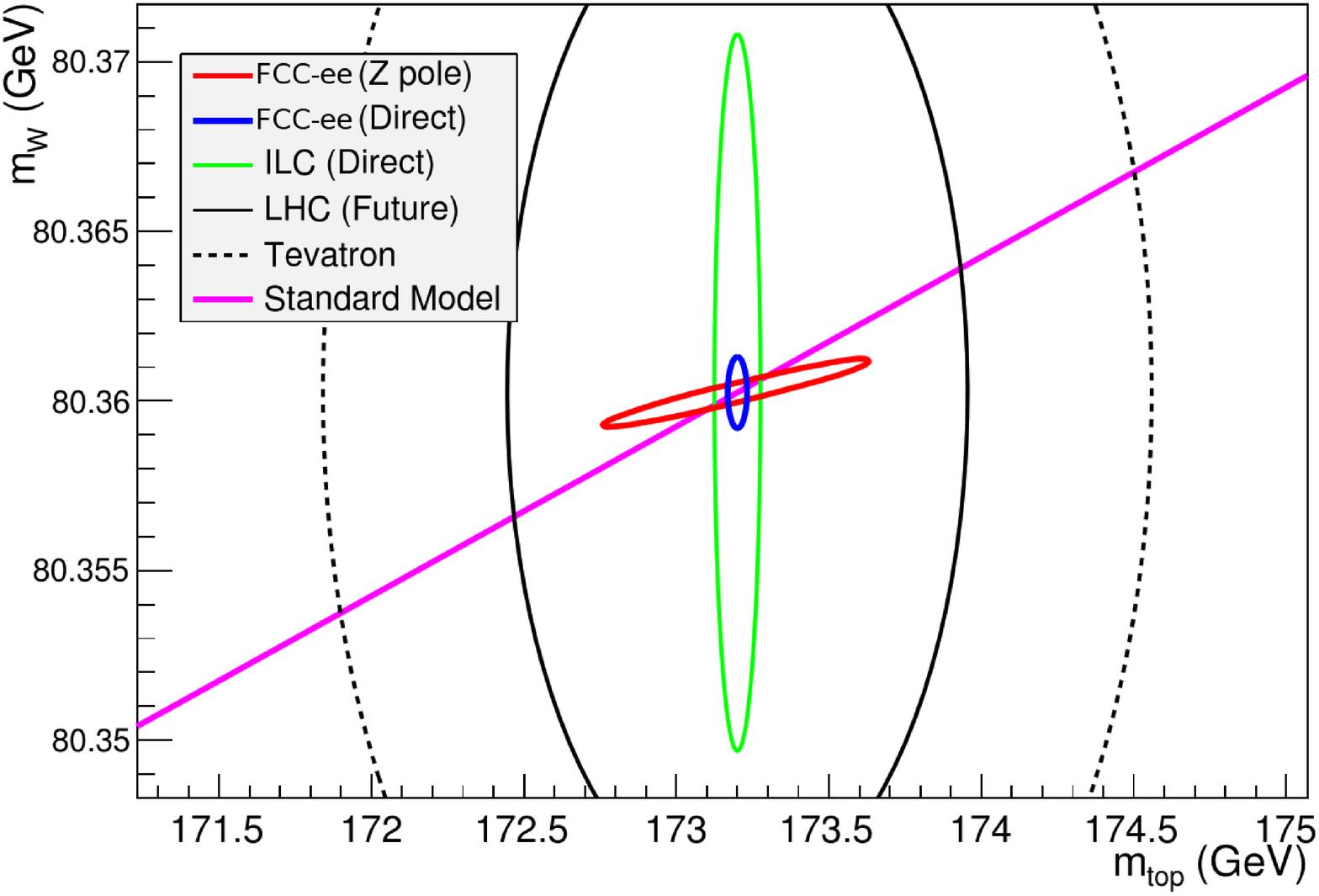}\hspace{0.2cm}
\includegraphics[width=0.45\columnwidth,height=4.7cm]{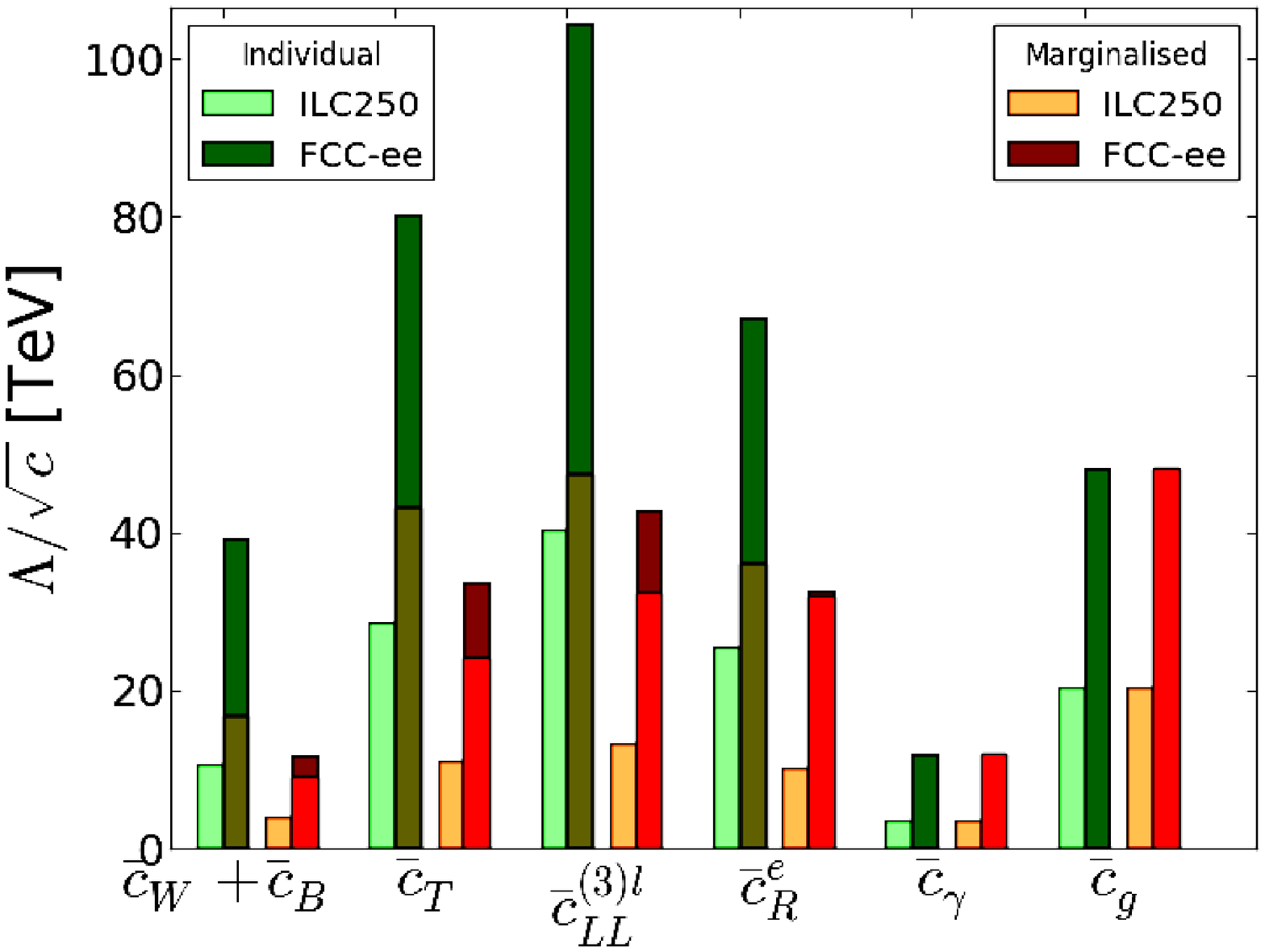}
\caption{Left: Comparisons of 68\% C.L. limits in the $m_\mathrm{t}$--$\mW$ plane at FCC-ee and other
  colliders~\cite{FCCee}. 
 Right: 95\% C.L. limits at FCC-ee(250~GeV) for leptonic operators sensitive to EWPT~\cite{Ellis:2015sca}.
}
\label{fig:FCCee_SM}
\end{figure}

None of these measurements can be carried out at the LHC (or other $\epem$ machines) with such a level of
precision. Physics beyond the SM can thereby be indirectly probed through loop corrections induced by possible new
heavy particles~\cite{Fan:2014vta}. Figure~\ref{fig:FCCee_SM} shows limits on 
the W-mass vs. top-mass plane (left), and on the subset of leptonic dimension-6 operators of a
model-independent Effective Field Theory of the SM parametrizing possible new physics (right)~\cite{Ellis:2015sca}.
Such measurements impose unrivaled constraints on new weakly-coupled physics. Whereas electroweak precision
tests (EWPT) at LEP bound any BSM physics at scales $\Lambda\gtrsim$~7~TeV, FCC-ee would reach up to
$\Lambda\approx$~100~TeV for some operators.

\section{Indirect  constraints on BSM via high-precision Higgs physics}
\label{sec:Higgs}

\begin{figure}[htpb!]
\centering
\includegraphics[height=4.75cm]{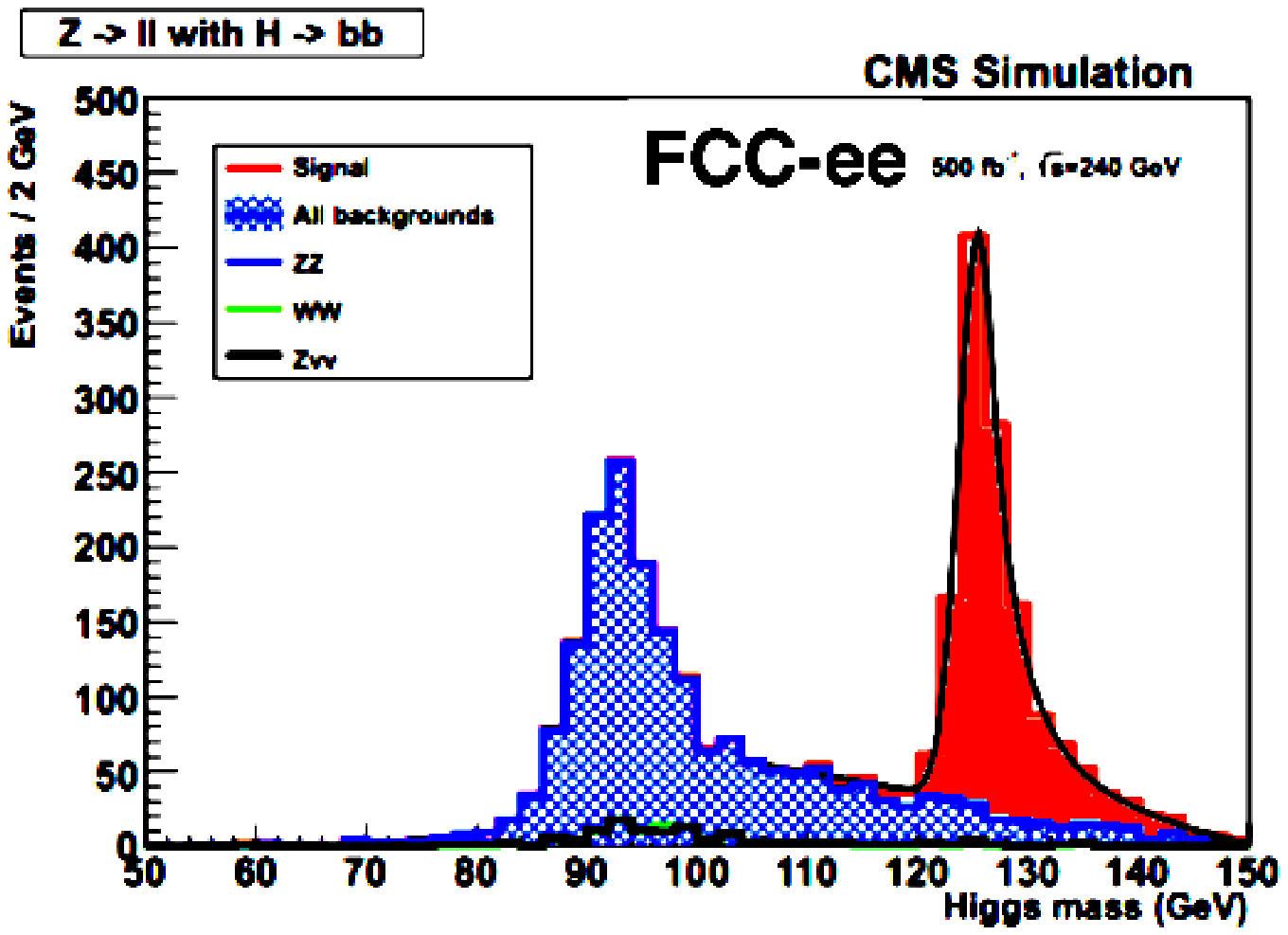}
\includegraphics[height=4.6cm,width=7.3cm]{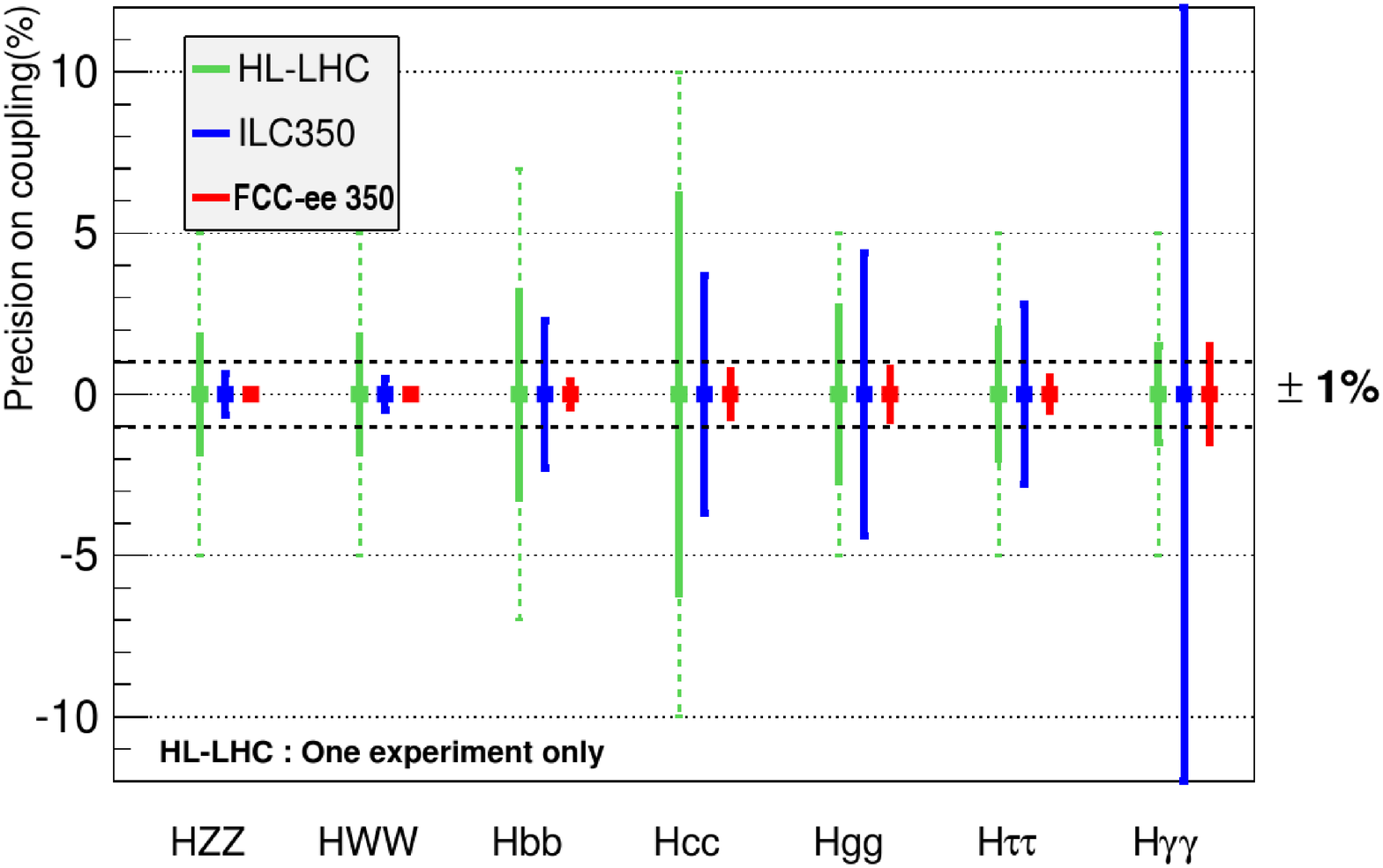}
  \caption{Left: Distribution of recoil mass against $\rm Z\to\ell\ell$ in the 
$\epem\to\rm HZ$ process with H$\to\bbbar$. Right: Comparison of expected relative uncertainties for the Higgs
boson couplings at future facilities~\cite{FCCee}.
\label{fig:Higgs}}%
\end{figure}

The Higgs sector of the SM can be probed with a unique precision with a high-luminosity lepton collider.
In the range of FCC-ee energies, Higgs production peaks at $\sqrts\approx$~240~GeV 
dominated by Higgsstrahlung ($\epem\to{\rm HZ}$), with some sensitivity at $\sqrts$~=~350~GeV also to
vector-boson-fusion ($VV\to {\rm H}\;\epem,\nu\nu$) and the top Yukawa coupling $y_{\rm t}$ (via
$\epem\to\ttbar$ with a virtual Higgs exchanged among the top quarks). The target total
number of Higgs produced at the FCC-ee (4 IPs combined, all years) amounts to 2.1~million at 240~GeV, 75\,000
in $VV\to{\rm H}$ at 350~GeV, and 19\,000 in s-channel $\epem\to{\rm H}$ at $\sqrts$~=~125~GeV
(Table~\ref{tab:runs}). Unique Higgs physics topics are open to study with such large data samples:
\begin{itemize}
\item High-precision model-independent determination of the Higgs couplings, total width, and exotic 
  and invisible decays (Fig.~\ref{fig:Higgs} right)~\cite{FCCee}.
\item Higgs self-coupling through loop corrections in HZ production~\cite{McCullough:2013rea}.
\item First-generation fermion couplings: (u,d,s) through exclusive decays $\rm H \to V\gamma$
  ($\rm V=\rho,\omega,\phi$)~\cite{Kagan:2014ila}, and electron Yukawa through resonant $\epem\to\rm H$ at 
$\sqrts=m_{_{\rm H}}$~\cite{DdE}. 
\end{itemize}
The recoil mass method in $\epem\to\rm HZ$ is unique to lepton colliders, and allows for an accurate tagging of
Higgs events (Fig.~\ref{fig:Higgs}, left) irrespective of their decay mode. It provides, in particular, a
high-precision ($\pm$0.4\%) measurement of $\sigma_{\rm \epem\to HZ}$ and, therefore, of $g_{\rm HZ}^2$. From
the measured value of $\sigma_{\rm \epem\to H(XX)Z}\propto\Gamma_{\rm H\to XX}$ and the different known decays
fractions $\Gamma_{\rm H \to XX}$, one can then obtain the total Higgs boson width with $\cO{1\%}$ uncertainty
combining different final states. The $\rm H\,Z(\ell^+\ell^-)$ final state can be used to directly measure the
invisible decay width of the Higgs boson in events where its decay products escape undetected, by analyzing
the distribution of the mass recoiling against the lepton pair. 
The Higgs boson invisible branching fraction can be measured with an absolute precision of 0.2\%. If not
observed, a 0.5\% upper limit (95\% C.L.) can be set on this branching ratio~\cite{FCCee}.

In addition, loop corrections to the Higgsstrahlung cross sections at different center-of-mass energies are
sensitive to the Higgs self-coupling. The effect is tiny but visible at FCC-ee thanks
to the extreme precision achievable on the $g_{\rm HZ}$ coupling. Indirect (model-dependent) limits on the
trilinear $g_{\rm HHH}$ can be set with a $\cO{70\%}$ uncertainty, comparable to that expected at HL-LHC~\cite{McCullough:2013rea}.
The large Higgs data samples available also open up the study of exotic (e.g. flavour-violating Higgs)
and very rare SM decays. The Higgs couplings to first- and second-generation fermions, which may reveal new
dynamics on the flavour structure of the SM~\cite{Ghosh:2015gpa}, can be accessed via exclusive decays $\rm
H\to V\gamma$, for $V=\rho,\omega,\phi$, with sensitivity to the u,d,s quark Yukawas~\cite{Kagan:2014ila}. 
The $\rm H\to\rho\gamma$ channel appears the most promising with $\cO{50}$ events 
expected. The low mass of the electron translates into a tiny $\rm H\to\epem$ branching ratio ${\rm BR}_{\rm
  \epem} = 5\cdot 10^{-9}$ which precludes any experimental observation of this decay mode and, thereby a
determination of the electron Yukawa coupling. The resonant s-channel production, despite its small cross
section~\cite{Jadach:2015cwa}, is not completely hopeless and preliminary studies indicate that could be
observed at the 3$\sigma$-level with 90~ab$^{-1}$ at $\sqrts$~=~125~GeV with a \cm energy spread commensurate
with the Higgs boson width itself ($\approx$4~MeV, requiring beam monochromatization)~\cite{DdE}.\\

In summary, the FCC-ee provides the smallest uncertainties for the measurements of Higgs boson couplings to gauge
bosons and fermions (Fig.~\ref{fig:Higgs}, right). Since any deviation $\delta g_{_{\rm HXX}}$, relative to the
SM value $g_{_{\rm HXX}}^{_{\rm SM}}$, can be approximately translated into BSM scale limits through the expression: 
$\rm \small \Lambda\gtrsim (1\,TeV)/\sqrt{(\delta g_{_{\rm HXX}}/g_{_{\rm HXX}}^{_{\rm SM}})/5\%}$. The
expected 0.15\% precision for the most precise coupling, $g_{\rm HZZ}$, would thus set competitive bounds, 
$\Lambda\gtrsim$~7~TeV, on any new physics coupled to the scalar sector of the SM. 

\section{Indirect constraints on supersymmetry}

Supersymmetry has many appealing features as a candidate framework for BSM physics, solving the naturalness
problem (through light stop contributions to the running of the Higgs boson mass), stabilizing the electroweak
vacuum, providing dark matter candidate(s) (the stable lightest SUSY particle, if R-parity is conserved),
predicting grand unification of gauge interactions, and connecting to string theory where SUSY is
preserved.
\begin{figure}[hbtp!]
\centering
\includegraphics[height=8.cm]{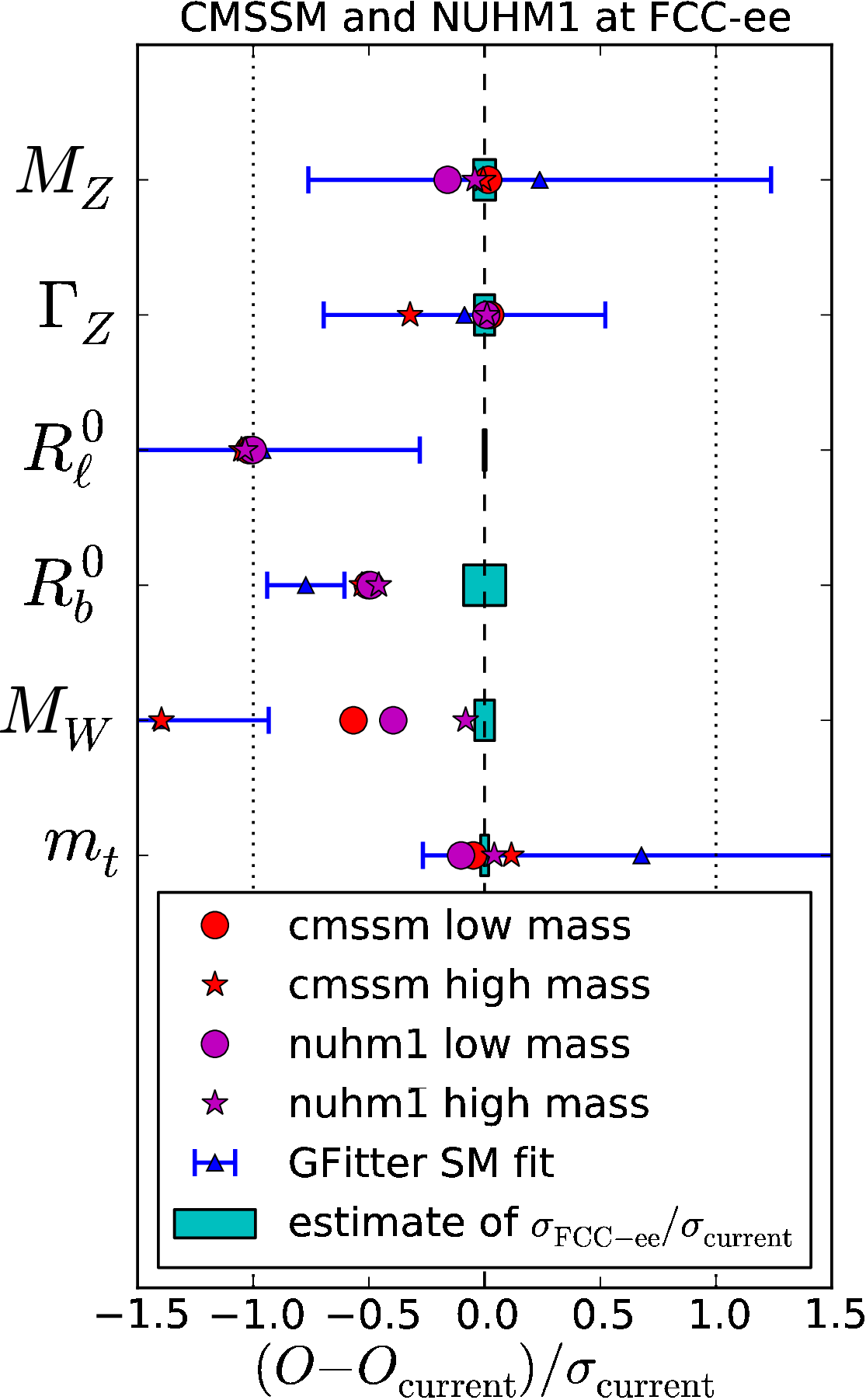} 
\includegraphics[height=8.cm]{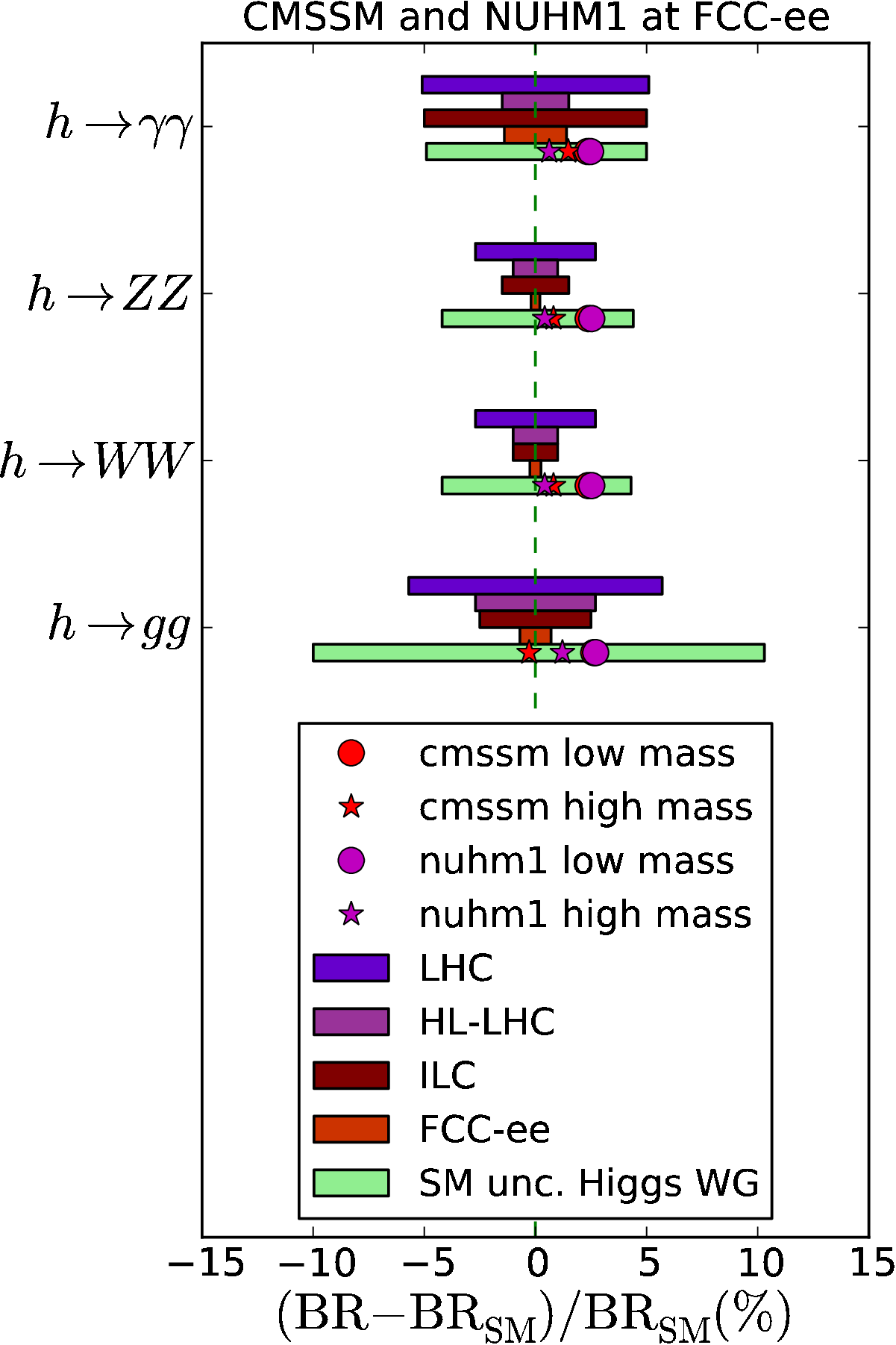}
\caption{Comparison of the estimated precision in EWPT (left) and Higgs branching ratios (right) measurements: 
  Current results (error bars, left; green boxes, right), low- and high-mass best-fit CMSSM and NUHM1 points 
  (circles and stars) and prospects at future colliders (bars)~\cite{Buchmueller:2015uqa}.
\label{fig:susy}}
\end{figure}
Since stops and higgsinos carry electroweak quantum numbers, $\epem$ colliders have better
sensitivity to those than to coloured sparticles (squarks, gluinos). The precision electroweak and top
(Section~\ref{sec:SM}) and Higgs (Section~\ref{sec:Higgs}) studies at the FCC-ee not only impose generic
constraints on new physics scales at  multi-TeV energies, but can also be interpreted in terms of sensitivity
to broad classes of SUSY models (such as the Constrained MSSM, Non-Universal Higgs Masses, or natural
SUSY)~\cite{Buchmueller:2015uqa,Fan:2014axa}. Figure~\ref{fig:susy} compares the precision on EWPT
(left) and Higgs (right) observables at the LHC, ILC and FCC-ee with the deviations from their SM values
expected for the low- and high-mass CMSSM and NUHM1 best-fit points~\cite{FCCee,Buchmueller:2015uqa}. 
It is clear that the FCC-ee has the best ability to distinguish these models from the Standard Model.
In particular, the high-mass CMSSM points in Fig.~\ref{fig:susy} provide an example of SUSY model which
likely lies beyond the LHC reach, featuring narrow strips where stop-neutralino coannihilation is important,
or focus-point strips at higher values of the ratio $m_0/m_{1/2}$. The EWPT and Higgs measurements at the
FCC-ee would be able to probe both types of narrow parameter-space strips that extend to large sparticle
masses, and indirectly determine CMSSM parameters also in such a pessimistic scenario~\cite{Buchmueller:2015uqa}.

\section{Direct constraints on BSM physics: dark matter and right-handed neutrinos}

\begin{figure}[htbp!]
\centering
\includegraphics[width=0.48\columnwidth,height=5.cm]{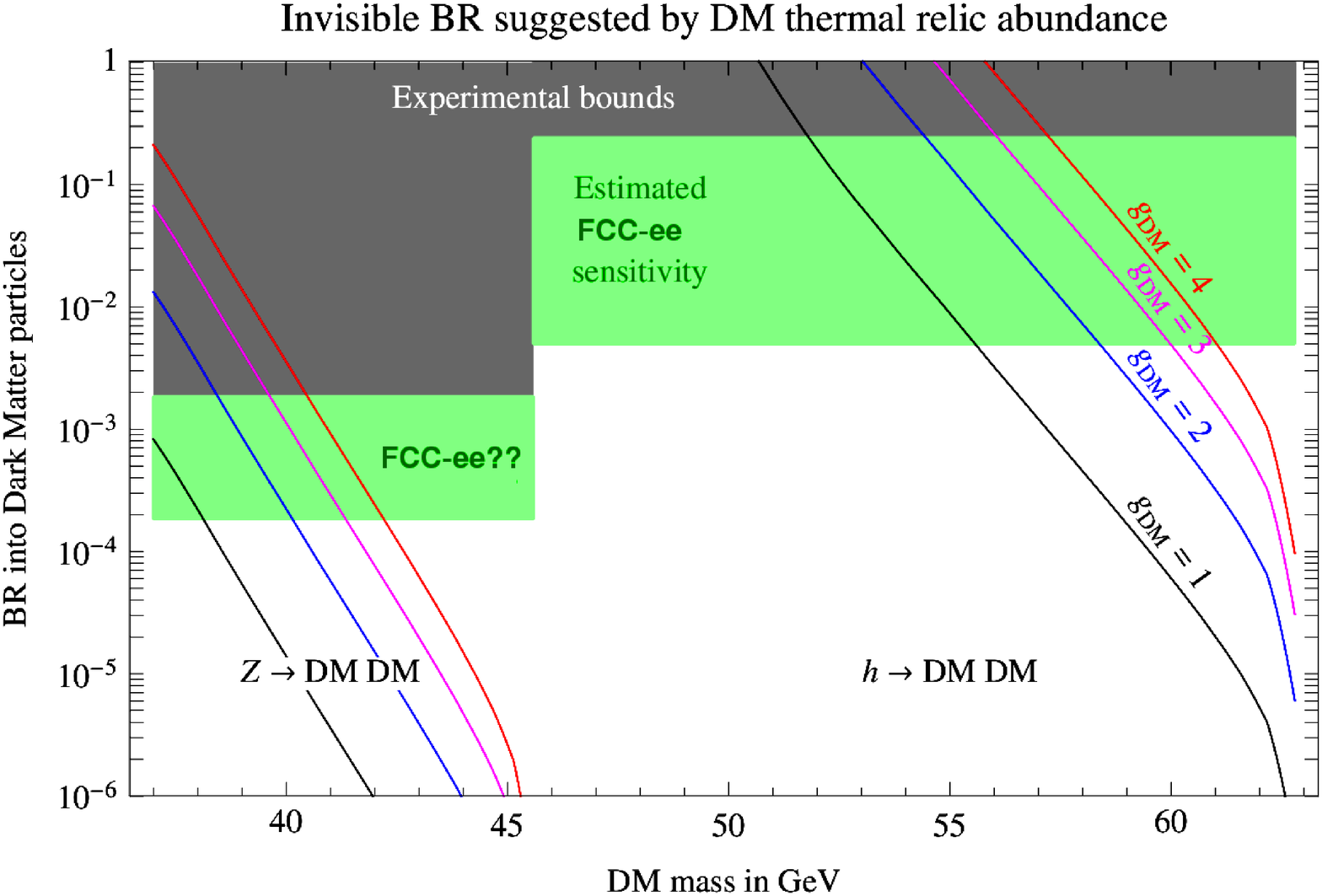}\hspace{0.1cm}
\includegraphics[width=0.49\columnwidth,height=5.cm]{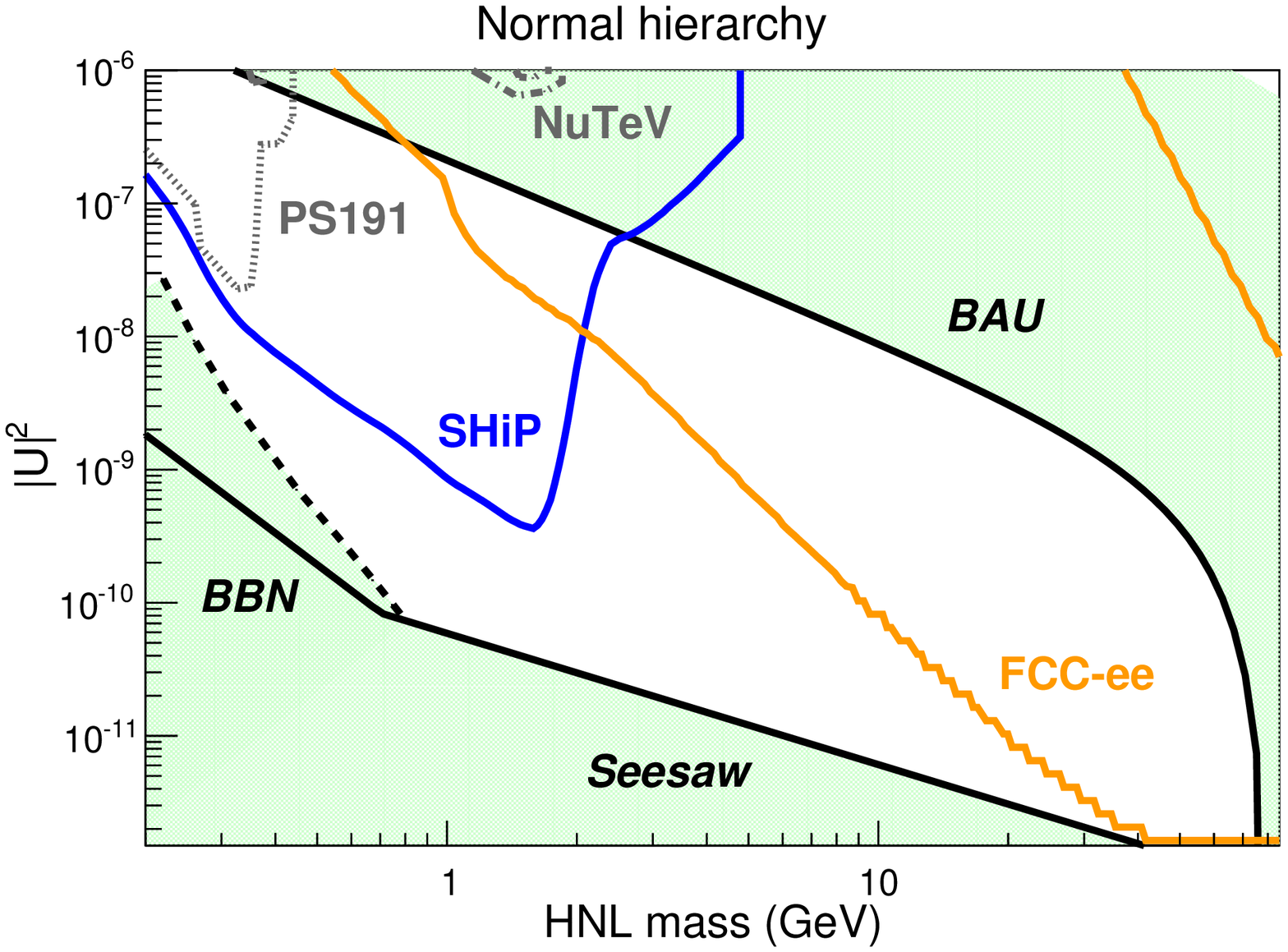}
\caption{Regions of FCC-ee sensitivity for: (i) Rare Z and H decays into DM pairs in the 
$\rm BR_{_{Z,H\to\,DM\,DM}}$ vs. $\rm m_{_{DM}}$ plane (left)~\cite{deSimone:2014pda}, and 
(ii) sterile neutrinos as a function of their mass and mixing to light neutrinos (normal hierarchy) for
10$^{13}$ Z decays (right)~\cite{Blondel:2014bra}.}
\label{fig:FCCee_BSM}
\end{figure}

The impact of the FCC-ee goes beyond indirect BSM studies and 
has also a strong discovery potential in direct searches of other key BSM extensions such as dark matter
(DM)~\cite{deSimone:2014pda} and right-handed neutrinos~\cite{Blondel:2014bra}, by exploiting the possibility to
measure very rare decays of the Z and H bosons into such new particles. As a matter of fact, measurements of
the invisible Z and H widths at the FCC-ee provide the best collider options to test DM lighter than $m_{_{\rm
    Z,H}}/2$ that couples via SM mediators. 
Figure~\ref{fig:FCCee_BSM} (left) shows the limits in the plane (branching ratio, DM mass) for the decays $\rm
Z,H\to DM\,DM$. Similarly, Fig.~\ref{fig:FCCee_BSM} (right) shows the unrivaled sensitivity of searches for
high-mass sterile neutrinos via decays $\rm Z\to N\nu_i$ (with $\rm N\to W^*\ell,Z^*\nu_j$) as a function of
their mass and mixing to light neutrinos (normal hierarchy)~\cite{Blondel:2014bra}. 

\section{Summary}

The FCC-ee has a unique program of searches for new physics via high-precision studies of the W, Z, and Higgs
bosons, and the top quark, with uncertainties at the permil level or below thanks to the huge luminosities
$\cO{1\rm{-}100}$~ab$^{-1}$ (for 4 interaction points) and the exquisite control of the beam energy in the
range $\sqrts \approx$~90--350~GeV. By searching for tiny deviations with respect to the SM predictions in a
rich set of measurements, BSM physics scales as large as $\Lambda \approx 7,100$~TeV for new particles
coupling to the scalar and electroweak SM sectors respectively, can be indirectly probed. When interpreted in
terms of broad classes of SUSY models, such precision electroweak and Higgs observables provide indirect
tests of supersymmetry in regions which are beyond the LHC reach. Last but not least, the FCC-ee covers also
direct BSM searches, through very rare and invisible Higgs and Z bosons decays which provide the best collider
options to test dark matter and sterile neutrinos with masses up to $m_{_{\rm DM,HNL}}\approx$~60~GeV.  

\section*{References}


\end{document}